\documentclass[12pt, onecolumn, draftcls, conference]{IEEEtran}
\IEEEoverridecommandlockouts

\usepackage[utf8]{inputenc}
\usepackage{amssymb,amsmath}
\usepackage{float}
\usepackage{tikz}

\newcommand{\colb}{\textcolor{black}}

\usepackage{hyperref}

\def\BibTeX{{\rm B\kern-.05em{\sc i\kern-.025em b}\kern-.08em T\kern-.1667em\lower.7ex\hbox{E}\kern-.125emX}}

\title{Improving Random Access with NOMA in mMTC XL-MIMO}
\author{\IEEEauthorblockN{{Thiago Augusto Bruza Alves}}
\IEEEauthorblockA{\textit{Department of  Electrical Engineering} \\
\textit{State University of Londrina  (UEL).}\\
86057-970, Londrina, PR, Brazil. \\
\url{thiagobruza@hotmail.com}\vspace{-5mm}}
\and
\IEEEauthorblockN{ {Taufik Abrão}}
\IEEEauthorblockA{\textit{Department of  Electrical Engineering} \\
\textit{State University of Londrina  (UEL).}\\
86057-970, Londrina, PR, Brazil. \\
 \url{taufik@uel.br}\vspace{-5mm}}
\thanks{This work was supported in part by the CAPES (Financial Code 001) and the National Council for Scientific and Technological Development (CNPq) of Brazil under Grant  310681/2019-7.} 
}

\begin{document}

\maketitle

\begin{abstract}
\colb{The extra-large multiple-input multiple-output (XL-MIMO) architecture has been recognized as a technology for giving support for the massive MTC (mMTC), providing very high-data rates in high-user density scenarios. However, the large dimension of the array increases the Rayleigh distance ($d_{{\rm Rayl}}$), in addition to obstacles and scatters causing spatial non-stationarities and  distinct visibility regions (VRs) across the XL array extension.} We investigate the random access (RA) problem in crowded XL-MIMO scenarios; the proposed grant-based random access {(GB-RA)} protocol combining the advantage of non-orthogonal multiple access (NOMA) and strongest user collision resolutions in extra-large arrays (SUCRe-XL) named NOMA-XL can allow access of two or {three colliding users} in the same XL sub-array (SA) selecting the same pilot sequence. \colb{{The received signal processing in a SA basis} changes the $d_{{\rm Rayl}}$, enabling the far-field planar wavefront propagation {condition, while} improving the system performance.} The proposed NOMA-XL GB-RA protocol is able to provide a reduction in the number of attempts to access the \colb{mMTC} network while improving the average sum rate, as the number of SA increases.
\end{abstract}	
\smallskip
\begin{IEEEkeywords}
NOMA; Machine Type Communication; XL-MIMO; Random Access Protocols; Crowded scenarios; {Uniform Rectangular Array (URA)}.
\end{IEEEkeywords}

\section{Introduction}
\colb{In B5G, the base station (BS) needs to simultaneously support devices with a variety of capabilities and deployments as the number of active devices continues to increase as well as access requests for multiple services, {\it e.g.}, vehicles, sensors, mobiles, etc., and applications in 5G and B5G, such as massive machine-type communication (mMTC) and crowded mobile broadband (cMBB) \cite{Fallgren13}. Hence, to support massive connections a promising candidate for the next generation of multiple access techniques, non-orthogonal multiple access (NOMA) has the ability to serve in the same resource element (RE), more than one device. In comparison to standard orthogonal multiple access (OMA), the NOMA increases the system throughput, improves user fairness, reduces latency, and enables large connections \cite{Akbar2021}, sharing the same orthogonal resource (time and frequency) by coding superposition in transmitter side and successive interference cancellation (SIC) in receiver side \cite{Maraqa2020}.}

\colb{{A high number} of devices attempting to connect to the network at the same time is a medium access control problem; standard random access (RA) systems are incapable of handling such a huge number of requests \cite{Clazzer2019}, and pure RA methods such as ALOHA have serious performance limitations. The adoption of NOMA and SIC considerably enhances performance, allowing two or more users per time slot \cite{Jako2015,Silva22}.}

\colb{In (over-)crowded scenarios, the number of user connection attempts considerably outnumbers the number of available pilot sequences. As a result, establishing collision resolution techniques became crucial for enabling efficient communication. The strongest user collision resolution (SUCRe) protocol is a well-known decentralized grant-based random access (RA) protocol for crowded massive multiple-input multiple-output (MIMO) systems, which takes advantage of MIMO properties \cite{Emil2017}, giving preference to users with good channel conditions while harming edge users. 
Hence, the SUCRe protocol has undergone several evolutions; {\it e.g.}, in \cite{Huimei2017b} a graph-based pilot access (SUCR-GBPA) protocol is proposed, enabling all users who lost contention resolution to choose a new pilot at random. Moreover, improving SUCRe by adding NOMA, the work in \cite{Pereira2021}  
compares the  NOMA-RA protocol to the classical SUCRe scheme, allowing users who try to access the medium with the same pilot signal to resolve collisions by distinguishing them in the power domain and achieving superior sum rate performance with reduced average latency.}

\begin{figure}[htbp!]
\vspace{-4mm}
\centering
\includegraphics[width=.65\linewidth]{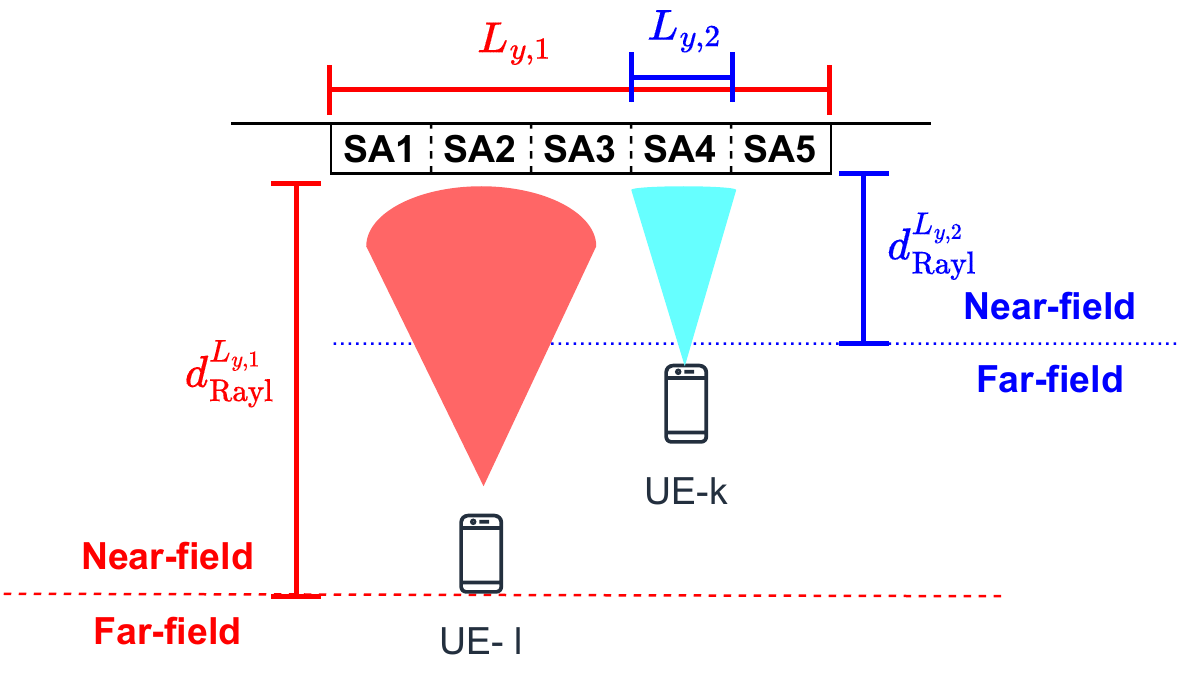}
\vspace{-3mm}
\caption{\colb{URA XL-MIMO with a $k$-th user, illustrating how a change in array aperture $D$ to $D'$ (subarray SA$_i$)  also changes  the Rayleigh distance $d_{{\rm Rayl}}$. In our random access XL-MIMO system, mMTC devices are distributed at random, located in a far-field area, {\it i.e.} $d_{\textsc{ue}_k} > d_{{\rm Rayl}}$.} }
\label{fig:ray_dist}
\end{figure}

\colb{The large-scale MIMO technology (XL-MIMO) distributes the antennas elements in a wide space, such as facades of buildings, shopping malls, or stadiums; therefore, typically, the devices, scatterers, and obstacles are located inside the Rayleigh distance of the {entire uniform rectangular array (URA) in Fig. \ref{fig:ray_dist}, which in this case is formed by five SAs}, resulting in different channel conditions, like {\it near-field propagation} and {\it spatial non-stationarity}. As a result, the received signal energy of a given UE varies significantly across the entire array; in addition, different regions of the array see different sets of scattered, obstacles, and UEs, based on the concept of visibility regions (VRs) \cite{Flordelis2020}.}

\colb{In this work, we adopt the far-field model for XL-MIMO scenarios where the UE-array  distance is typically higher than the Rayleigh distance, $d_{\textsc{ue}_k} > d_{{\rm Rayl}}$, since the received signal processing is accomplished in a SA basis, reducing the array aperture $D$. The  Rayleigh distance  is quadratically proportional to the array aperture $D$  as:}

\colb{
\begin{equation}\label{eq:dRay}
d_{\rm Rayl} = 2\frac{D^2}{\lambda_\textsc{c}}, 
\end{equation}
\noindent where
$\lambda_\textsc{c}$ is the {carrier} wavelength. Hence, splitting the entire array in SAs changes considerable the limit of near-field and far-field, by changing the array aperture to $D_{\textsc{sa}}=\frac{D}{B}$, where $B$ is the number of SAs. Therefore, users in the near-field area when considering the entire array can be seen on a SA basis as in far-field planar wave propagation.}

\colb{{The non-stationarity} can be exploited by applying the concept of VR, where an association with the geographic area of the cell and the visible portion of the array or SA, as demonstrated in \cite{Nishimura2020} where SUCRe was adapted to the XL-MIMO context. With that, a certain UE is present in a region and has an associated VR due to the associated scatterers. For simplicity, we consider that an entire SA can be visible to the device, {\it i.e.,} is part of its VR. Thus, the probability of the $b$-th SA being visible by the $k$-th UE can be exploited to improve the performance of the RA protocol in XL-MIMO.}

\colb{In \cite{Iimori2022}, a solution for the random access problem is addressed by jointly estimating the activity and the channels of UEs in XL-MIMO subjecting to non-stationarity in the far-field propagation scenarios; the effectiveness of the proposed algorithm is demonstrated in terms of the {\it normalized mean square error} (NMSE) and {\it activity error rate} (AER) performance indicators.} 
\colb{Differently from \cite{Alves2022} and \cite{Bruza2023}, in this work, by optimizing the scale factor $\delta^*_{\text{NOMA}}$, we admit more than two users (up to three users) colliding by the same pilot sequence under NOMA scheme, maximizing the sum-rate for different loading users and the number of SAs. Notice that herein, we treat and process the received signal at SA level; hence, by dividing the array into SA one can manipulate the Rayleigh distance $d_{\rm Rayl
}$, so users still experience the far-field condition.}

\colb{The \textit{Contribution} of this paper is twofold: 
\textbf{a}) a grant-based RA protocol is proposed, based on power-domain NOMA and by exploring the overlapping VR of XL-MIMO systems (NOMA-XL), which combines the benefits of XL-MIMO systems and NOMA, allowing until three of users selecting the same pilot sequence in at least one SA.
\textbf{b}) the composition of the array into SAs, change the Rayleigh distance, allowing UEs to stay in the far-field area; in this condition, numerical analyses are carried out taking into account the superimposed received signal processed in an SA basis, the optimal scale factor $\delta^*_{\text{NOMA}}$, and performing SIC to identify the signal of each user.   
}

\section{System Model} \label{System_model}

\colb{A time-division-duplexing (TDD) system is considered, in which a BS deploys an extra-large  URA placed on} the $y$-$z$ plane with a total of $M$ antenna elements, i.e., $M_y \times M_z$, where $M_y$ and $M_z$ denote the number of antenna elements along the $y$- and $z$-axis, respectively, with the firth line of \colb{the array} starting at the origin $O$ and deployed along the ordinate axis.
\colb{In the same way, as} \cite{Nishimura2020, Marinello2022}, we consider a simplified bipartite graph model in XL-MIMO, the array is divided into $B$ subarrays (SAs), each composed by a fixed number of $M_b = M/B$ antennas, ensuring the minimum antennas elements $\left (M_b \geq 50 \right)$.

\colb{UEs equipped with one antenna are randomly distributed in a square cell and distanced beyond the Rayleigh distance ($d_{{\rm Rayl}}$) limit from the array, and grouped in a set} $\mathcal{U}$, being $\mathcal{A} \subset \mathcal{U}$ the subset of active UEs, with \colb{temporarily} dedicated data pilots, and $\mathcal{K} = \mathcal{U} \setminus \mathcal{A}$ be the set of inactive UEs (iUEs).
\colb{The cardinality $|\mathcal{K}| = K$ represents the number of iUEs in the cell, and it has no dedicated pilot, in case the UE needs to become active, they must be assigned to one. The BS only allocates pilots to active devices and reclaims the pilots when needed. The iUEs try a RA attempt with probability $P_a$.}   

Let $\mathcal{M}$ denote the set of all BS SAs and $\mathcal{V}_k \subset \mathcal{M}$ the subset of SAs visible to the $k$-th UE. To compose the mathematical modeling, we assume that the subset $\mathcal{V}_k$ is generated randomly, where each SA is visible with probability $P_b$ by UE $k$, like in \cite{Nishimura2020,Marinello2022}. This probability simulates the influence of random obstacles and scatterers in the environment interacting with the signals transmitted by/to the UEs, resulting in VRs. \colb{We assume that each UE previously knows its VR $\mathcal{V}_k$, by deploying the same measurement procedure as in \cite{Carvalho2020}.}

We define $\tau_{\text{RA}}$ as the number of RA mutually orthogonal pilot sequences that are possible $\textbf{s}_1,\dots,\textbf{s}_{\tau_{\text{RA}}} \in \mathbb{C}^{\tau_{\text{RA}} \times} 1$, each with length $\tau_{\text{RA}}$, and $\left \| \textbf{s}_{\tau_{\text{RA}}} \right \| = \sqrt{\tau_{\text{RA}}}$.

\begin{figure}[htbp!]
\centering
\includegraphics[width=.85\linewidth]{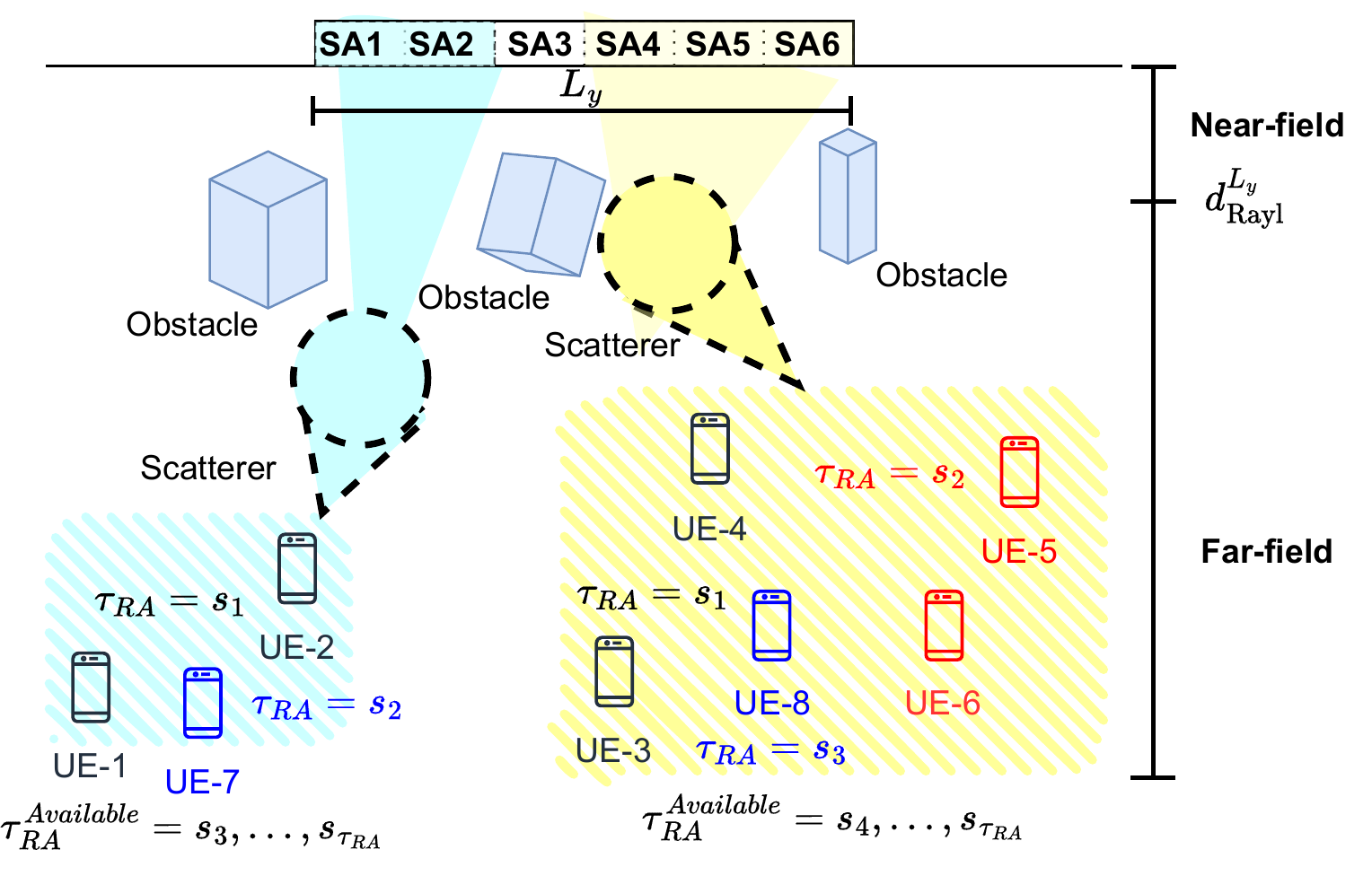}
\vspace{-8mm}
\caption{Crowded MTC system with URA divided into a 6-SA; the colored regions in each SA define visible UEs in far-field area, $P_b$ is visible success probability.}
\label{fig:distance}
\end{figure}

Since each SA visibility follows a Bernoulli distribution with success probability $P_b$, it is crucial to note that when an antenna is visible to a user, it indicates that the user may transmit/receive signals to/from this antenna with a non-zero channel gain \cite{Marinello2022}.
\colb{We assume an overcrowded scenario where $K$ iUEs $\left (K \gg M \right)$ are randomly distributed within a square cell in front of the XL-MIMO entire array.  Modifications in the array aperture, {\it e.g.} the SA, impact the Rayleigh distance, since the array aperture scale down from $D=L_y$ to $D_{\textsc{sa}}=\frac{L_y}{B}$. As a result, in terms of SA level, the XL-MIMO channel can be modeled as a far-field propagation considering typical application scenarios, as illustrated in Fig. \ref{fig:distance}.}

\subsection{Channel Model}
\colb{We model the NLOS channel propagation, and the location of the elements follows a 3D representation \cite{Lu2022}, ensuring the mMTC devices are physically far away from the Rayleigh distance.} The physical dimension of the URA along the $y$- and $z$-axis are $L_y = (M_y)d_m$ and $L_z = (M_z)d_m$, where $M_y$ and $M_z$ denotes the number of antenna elements along $y$- and $z$-axis, and $d_m$ is the distance between elements. \colb{We adopt the far-field planar wavefront propagation model based on the Rayleigh distance, which is directly linked to the array aperture as defined in eq. \eqref{eq:dRay}.
Thus, we adopted the far-field model where} the location of device $k$ is $\textbf{q}_k$ and the location of each URA antenna element $m_{z,y}$ is $\textbf{w}_{m_{z,y}}$, hence, the distance of the $k$-th UE to the specific antenna element $m_{z,y}$ is determined by  \cite{Lu2022}:
\begin{equation}
r_{k,m_{z,y}} = \left \| \textbf{w}_{m_{z,y}} - \textbf{q}_k \right \| {> d_{\rm Rayl}}
\end{equation}
 
Considering  an urban micro scenario, the large-scale fading model includes path loss and shadowing \cite{Fallgren13}: 
\begin{equation}
\beta_{k,m_z,m_y} = 10^{-\kappa \log(r_{k,m_z,m_y})+\frac{g+\varphi}{10}},
\label{eq:beta}
\end{equation}
\noindent where $g$ = $-$34.53 dB is the path-loss at the reference distance, the path-loss exponent $\kappa$ = 3.8, and $\varphi \sim \mathcal{N}(0,\sigma^2_{\text{sf}})$ is the shadow fading, a log-normal random variable with standard deviation $\sigma_{\text{sf}}$ = 10 dB. In same way as \cite{Nishimura2020,Marinello2022}, we adopt the average large scale fading for the $k$-th user at the $b$-th SA, $\beta_k^{(b)} =$ $\frac{1}{M_b} \sum_{m=1}^{M_b} \beta_{k,m}$. Let $\textbf{h}^{(b)}_k \in \mathbb{C}^{M_b \times 1}$ be the Rayleigh fading channel vector between UE $k \in \mathcal{K}$ and SA $b$, following $\textbf{h}_k^{(b)} \sim \mathcal{CN}$ $\left ( 0,\beta_k^{(b)} \textbf{I}_{M_b} \right )$.

\vspace{-3mm}

\vspace{-1mm}
\section{Random Access for NOMA XL-MIMO}
The {\bf SUCRe protocol} \cite{Emil2017} is a decentralized RA method for resolving pilot collisions on the user side. The protocol consists of following steps: $\textbf{a)}$ $K P_a$ users randomly choose an uplink (UL) RA pilot and transmit it; $\textbf{b)}$ the BS receives such UE signals and responds with precoded downlink (DL) pilots; $\textbf{c)}$ Using the received signal, UEs can estimate the total of the signal strengths of UEs competing for the same RA pilot chosen by them, compare the such estimate to its signal gain, and retransmit the chosen RA pilot if it judges it is the strongest UE among the contenders; $\textbf{d)}$ in conclusion, the BS allocates dedicated data pilots to the UEs not colliding in Step before. In mMTC scenarios, $K$ is large compared to the number of antennas $M$ $\left ( K \gg M \right )$, in contrast, $P_a \ll 1$.

\subsection{Exploring VRs, SAs and NOMA in NOMA-XL Protocol}
The proposed NOMA-XL protocol combines the \colb{improvement} of non-overlapping VRs of specific UEs along the SAs operating under the SUCRe-XL protocol \cite{Nishimura2020}. Besides, our proposed protocol can proceed with up to two or three users \colb{in overlapping VRs (in SA-based)} selecting the same pilot \colb{and physically in a far-field area.} Steps I and II are similar to the SUCRe protocol in  \cite{Emil2017}.

\noindent \textbf{Step I}: $KP_a$ UEs make an RA attempt. 

\noindent \textbf{Step II}: SAs send a precoded DL pilot to UEs from Step I. 

\noindent\textbf{Step III}: By exploring the power domain NOMA, users colliding pilots can be served at \colb{the} same resource (SA, frequency, time, pilot). 
Due to imperfections in the SIC process, in this work, the proposed protocol can accept collisions \colb{of up} to three users, {\it i.e.}, performing up to two SIC steps. However, in collisions between four or more users in at least one SA, the protocol could fail to resolve the collision, and unsuccessful UEs are instructed to try again on the next RA with probability $P_{\rm na}$, limited to a maximum of 10 attempts.

The collision resolution follows the decentralized decision rule, \textit{i.e.} each users decides whether to continue in access attempt; following the decision rule \colb{\cite{Nishimura2020}:} 
\begin{equation}
\rho_k \sum_{m \in \mathcal{V}_k} \beta_k^{(m)} \tau_{\text{RA}} \, > \, \frac{\colb{\widehat{\alpha}}_{t,k}}{2} + \epsilon_k,
\label{eq:decision}
\end{equation}
\colb{where $\widehat{\alpha}_{t,k}$ is the estimation of the sum of signal gains of contending UEs with its own $\rho_k \sum_{m \in \mathcal{V}_k} \beta_k^{(m)} \tau_{\text{RA}}$ by the same pilot $\tau_{\text{RA}}$, and $\epsilon_k$ is the bias term. If inequality in Eq. \eqref{eq:decision} holds, the $k$-th user repeats its pilot transmission and forms the new set $\mathcal{S}_t$; otherwise, it remains in silence and tries to communicate in the next RA stage with probability $P_{\rm na}$.
The {\it bias term} is given by:}
\begin{equation}
\epsilon_k = \frac{\delta^*_{\text{NOMA}}}{\sqrt{M_b} \cdot \sum_{b \in \mathcal{V}_k} \beta_k^{(b)}},
\end{equation} 
\noindent \colb{where $\delta^*_{\text{NOMA}}$ is the optimal \textit{scale factor} that should be adjusted to improve the average sum rate of the system.}

The main feature of our NOMA-XL RA protocol is the modification in the contention resolution process proposed in \cite{Nishimura2020}. The signal is processed in each SA to decode superimposed signals; hence, a SIC strategy is evoked to cancel and decode users' signals sharing the same pilot sequence.
Without loss of generality, in the contention resolution step, the protocol accepts up to \colb{three UEs} in \colb{the} NOMA cluster assuming power disparities on a SA basis; hence, the BS proceeds with the SIC according to the descending order of channel gains $(\left \| \textbf{h}_1 \right \| > \left \| \textbf{h}_2 \right \| > \left \| \textbf{h}_3 \right \|)$, {\it i.e.,} the strongest UE-1 signal is decoded first without SIC application, followed by UE-2 signal with one step of SIC to partial removed UE-1 signal and interference of UE-3 and in the sequel decodes the signal of UE3 after a two steps SIC to partially cancel signal of UE1 and UE2.

The signal-to-interference-plus-noise ratio (SINR) in each SA,
selecting the same pilot sequence $t$ is defined as \cite{Bruza2023}: 
\begin{align} 
\label{eq:SINR}
\gamma ^{(b)}_{t,k} = \frac {p_{k} \beta ^{(b)}_{k}}{\varpi_1 \sum ^{k-1}_{j=1}p_{j} \beta ^{(b)}_{j}+\sum ^{\left |{ \mathcal {S}_{t} }\right |}_{i=k+1}p_{i} \beta ^{(b)}_{i} + \sigma ^{2}}
\end{align}
where $\varpi_1$ is the residual interference factor after applying SIC, assuming imperfect signal reconstruction.

\noindent\textbf{Step IV}: \colb{We consider that the BS perfectly estimates the channel} to the user, or users, using the pilot signal sent in \colb{Step III}. After that, the corresponding message is decoded. If the decoding is successful, the BS has identified users in set $\mathcal{S}_t$ and admits it to the payload coherence blocks by allocating a dedicated data pilot sequence. If the decoding fails, the protocol also fails to resolve the collision. The unsuccessful UE is instructed to try again after a random interval, limited \colb{to a maximum} of 10 attempts. It stops sending the pilot sequence, and the packet is considered lost.

\subsection{System Sum-rate} 

\colb{To compare the RA protocols, we define the sum-rate metric}. In SUCRe-XL, only one user is accepted per pilot sequence in non-overlapping VR; on the other hand, in \colb{NOMA-XL}, there will also be two \colb{or three users} who consider themselves winners in the contender. The sum rate of NOMA-XL in SA basis can be defined using \colb{Eq. \eqref{eq:SINR} as}:
\begin{equation}
R^{\text{NOMA-XL}}_{\sum} = \sum_{b=1}^B \sum^{\tau_{\text{RA}}}_{t=1}  \sum_{k=1}^{\left | \mathcal{S}_t \right |} \log_2 \left (1+ \gamma^{(b)}_{t,k} \right),  \quad [\rm bpcu]
\end{equation}
 The sum rate of SUCRe-XL can be defined as: 
\begin{equation}
    R_{\sum}^{\text{SUCRe-XL}} = \sum_{b \in \mathcal{V}_k} \sum^{\tau_{\text{RA}}}_{t=1} \log_2 \left ( 1 + \frac{\rho_k \beta_k^{(b)}}{\sigma^2} \right ), \quad [\rm bpcu]
\end{equation}
\noindent assuming that just one user per pilot sequence is accepted, and \colb{$\delta^{(K,B)^*}_{\text{SUCRe-XL}}=-1$}.

\vspace{-1mm}

\section{Numerical Results}

It is assumed a \colb{$L_y=10$ meters} URA with $M = 500$ antennas in a \colb{100} x 100 $\rm m^2$ cell with $K$ inactive users (iUES) randomly distributed \colb{in far-field area of the cell, and the number} of available pilots is $\tau_{\text{RA}} = 10$ sequences. The normalized transmit power $\rho_k = 1 \rm W$, other simulation parameters are listed in Table \ref{tab:simulations}.

\subsection{System sum rate} 
The optimal value of \textit{scale factor}, $\delta^{(K,B)^*}_{\text{NOMA-XL}}$ is obtained numerically, by exhaustive search approach, \colb{as in \cite{Pereira2021}}; so, it is optimal in the sense of sum rate maximization; hence, each $K$-iUE and $B$-SA configuration results in an optimal parameter.

\begin{figure}[!htbp]
\vspace{-3mm}
\centering
\includegraphics[width=.81\linewidth]{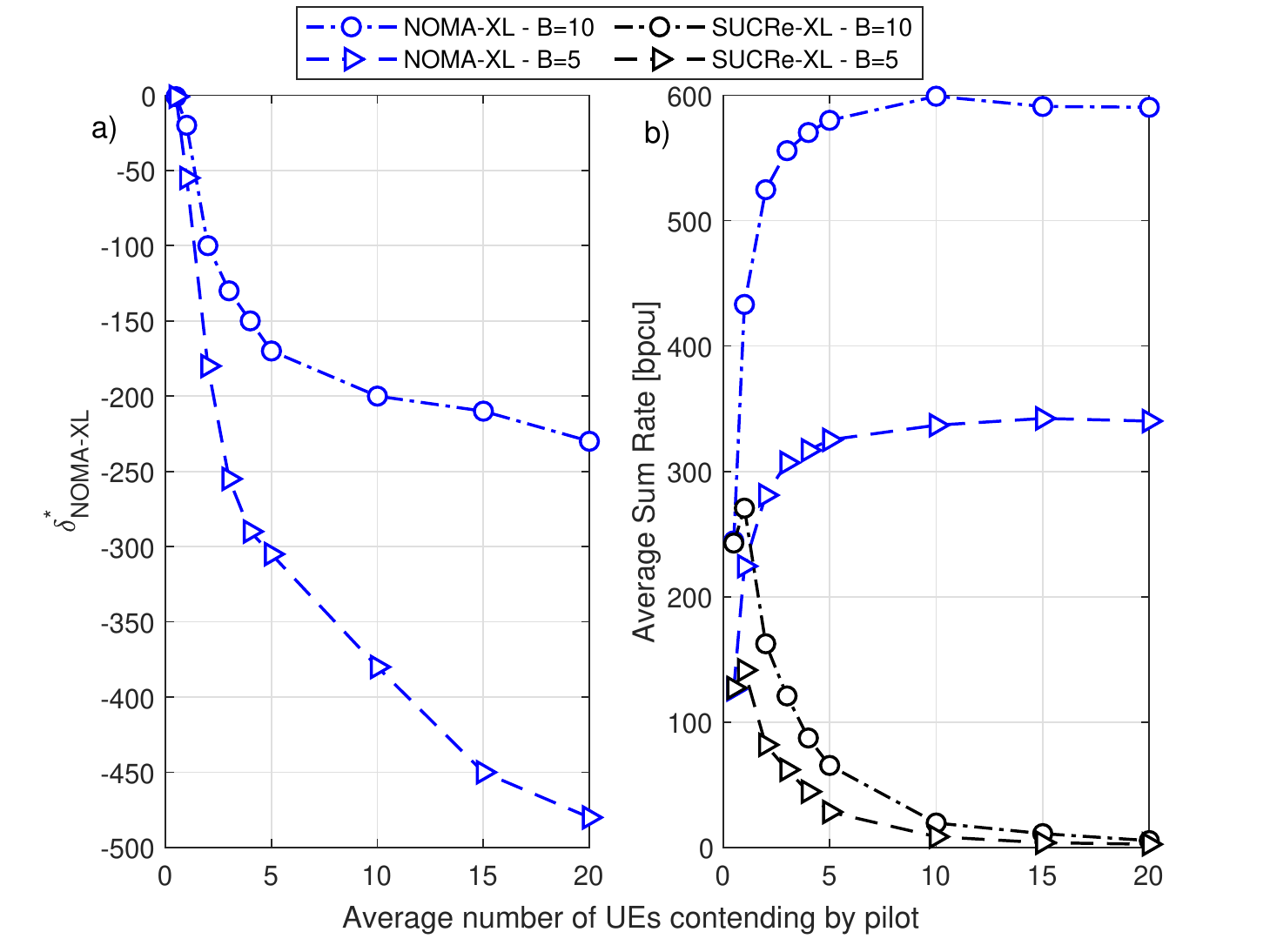}
\vspace{-6mm}
\caption{\colb{ {\bf a}) Optimal {\it scale factor} for average sum-rate maximization for each $\{K; B\}$ scenario, obtained via exhaustive search. {\bf b}) Average maximum sum-rate in each $K$ and $B$ scenario, attained after adjusting $\delta^*_{\text{NOMA-XL}}$.}} 
\label{fig:delta}
\end{figure}

\colb{Fig. \ref{fig:delta} a)} shows the optimal \textit{scale factor} parameter obtained numerically \colb{for each scenario of $K$ and $B$ by exhaustive search approach} and it correspondent maximum NOMA-XL \colb{sum rate in Fig. \ref{fig:delta} b)}, 
\colb{notice} that the baseline (SUCRe-XL) requires the non-overlapping VR em each SA to accept the device; hence, presents a modest gain even increasing the number of SA. The highest sum rate values occur before the saturation of usage of available pilots. On the other hand, the NOMA-XL processes the signal superimposed in each SA, canceling the more robust user's signal, with a SIC residual interference assumed in order of $\varpi_1 = 0.1$, and achieving improved results when $B$ increases. Also, a maximum sum rate limit is reached when adjusting the {\it scale factor} $\delta$, limiting the maximum number of users  selecting the same pilot to three users, when the number of iUEs grows the \textit{scale factor} decreases. This decay can be observed when the number of SA is lower, corroborating that the greater the interference of users, the smaller the $\delta$ factor.

\begin{table}[!ht]
\caption{Simulation Parameters.}
\centering
\begin{tabular}{l|c}
\hline \hline
\textbf{Parameter}  & \textbf{Value}      \\ 
\hline \hline
Antennas elements y-axys, z-axys & $M_y = 100, M_z = 5 $ \\ \hline
Number of BS antennas (URA)  & $M = M_y \times M_z = 500$     \\ \hline
\colb{Antenna element distance} & \colb{$d_m = 0.1$m} \\ \hline
\colb{Array aperture}   & $D=M_y \times d_m$ \\ \hline 
User transmit power       &  $\rho_k =1 \rm W$     \\ \hline
Noise Power     & $\sigma^2 = 1 \rm W$ \\ \hline
Number of iUEs               & $K \in {[}0;15000{]}$ \\ \hline
\colb{Carrier wavelength} & \colb{$\lambda_\textsc{c} = 0.125$m} \\ \hline
Access probability ($1^{st}$ attempt) & $P_a = 0.01$ \\ \hline
Access probability (new attempts)       & $P_{na} = 0.5$    \\ \hline
\colb{Success probability of each SA being visible} & \colb{$P_b = 0.5$} \\ \hline
Number of available pilot sequences     & $\tau_{\text{RA}} = 10$ \\ \hline
SIC imperfection factor                 & $\varpi_1= 0.1$     \\ \hline
Monte Carlo realizations     & 5000 \\ \hline \hline
\textbf{\# SAs  \hspace{6mm} Sub-Array Aperture, $D_\textsc{sa}$} & \textbf{Rayleigh dist., $d_{\rm Rayl}$} \\ \hline 
$B = 1$ \hspace{18mm} $10$m    & $1600$m         \\ \hline
$B = 5$ \hspace{19mm} $2$m     & $64$m  \\ \hline
$B = 10$ \hspace{17.6mm} $1$m   & $16$m         \\ \hline \hline
\end{tabular}
\label{tab:simulations}
\end{table}

\subsection{Average Number of Access Attempts} 
\begin{figure}[htbp!]
\centering
\includegraphics[width=.81\linewidth]{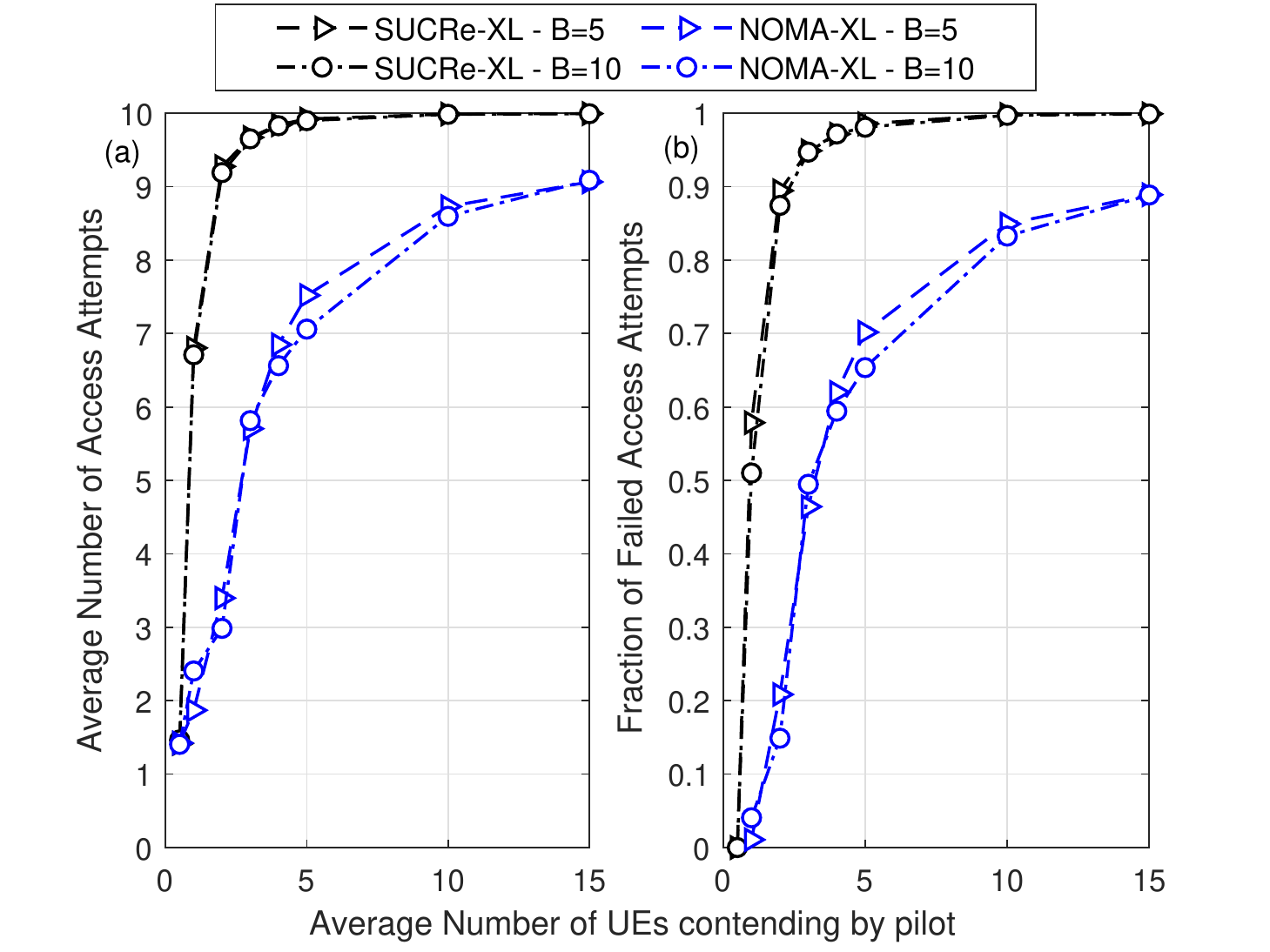}
\vspace{-5mm}
\caption{RA systems performance. The Baseline SUCRe-XL and NOMA-XL. (a)  Avg. Number of RA Attempts. (b) Prob. of failed access attempts.}
\vspace{-2.5mm}
\label{fig:ANAA}
\end{figure}

Fig. \ref{fig:ANAA} shows that the proposed protocol NOMA-XL in crowded scenarios $(1000 < K < 15000)$  achieves improved RA performance. The modification in decentralized collision resolution is that NOMA-XL accepts two or three users colliding on the overlapping VR, thanks to the power diversity; therefore, the superposition signal can be processed after SIC process in each SA. With the increase in the number of SAs $B$, but constrained by the channel hardening and favorable propagation properties, the probability of collision in each subarray by sharing overlapping VRs in each SA increases. Such collision resolution configuration is impossible to treat with SUCRe-XL due to the absence of the interference cancellation step. As a result, the NOMA-XL allows decreasing the number of access attempts even under solid interference in the same SA.

\begin{figure}[!htbp]
\centering
\includegraphics[width=.81\linewidth]{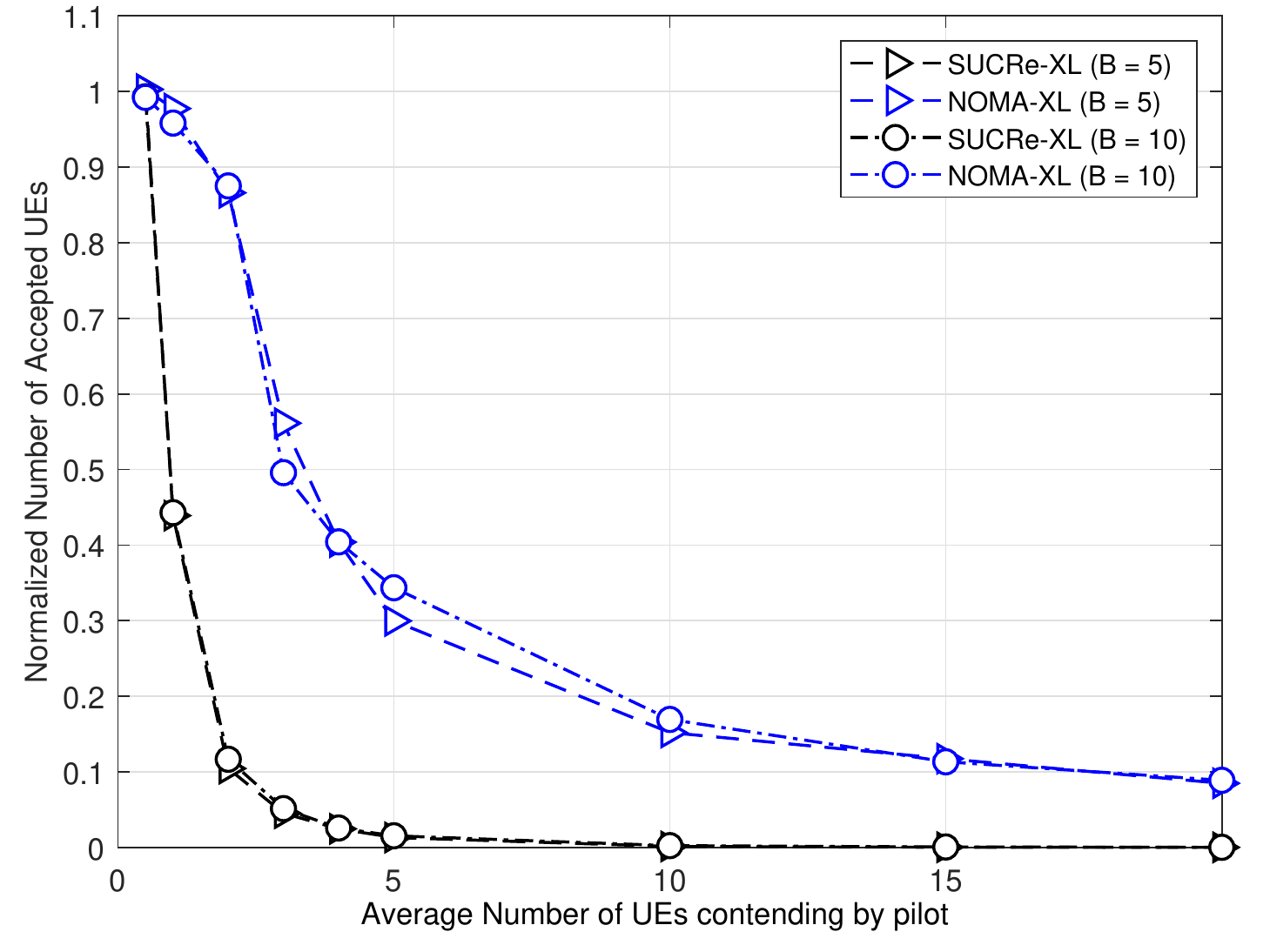} \vspace{-4mm}
\caption{Normalized number of accepted UEs, obtained with $\delta^*_{\text{NOMA-XL}}$.}
\label{fig:extra}
\end{figure}

The normalized number of accepted UEs, calculated by the number of users accepted by the number of users trying to access, given in Fig. \ref{fig:extra} corroborate with the improved results in NOMA-XL to accept UEs to transmit due to the superposition signal can be processed by SIC in each SA.  

\section{Conclusion}
By exploiting NOMA and signal processing in each SA the grant-based RA protocol performance can be improved. Indeed, the XL-MIMO adds a new degree of freedom with the received signal being processed by different SAs, together with the degree of freedom introduced by power domain NOMA, allowing to resolve \colb{of} the collision until three users using the same pilot sequence in the same VR subarray, and by applying the SIC stage in each SA at the BS side. \colb{We demonstrate that by processing the received signal on a SA basis can change the Rayleigh distance and place users in the far-field area.} Anchored in our numerical results, we found a substantial \colb{improvement} of the proposed RA protocol regarding the recent literature SUCRe-XL method, achieving both a reduction in the number of access attempts and a significant increase in the sum rate under crowded scenarios analyzed $(1000<K<15000)$.


\end{document}